\documentclass[a4paper,10pt,twocolumn,groupedaddress,aps,prl, showpacs]{revtex4-1}
\usepackage[utf8]{inputenc}
\usepackage{mathtools}
\usepackage[pdftex]{graphicx}
\usepackage{epstopdf}
\usepackage{amsmath}
\usepackage{amssymb}
\usepackage[left=1.5cm,top=2.2cm,right=1.5cm,bottom=2.2cm]{geometry}
\usepackage{relsize}
\usepackage{siunitx}
\usepackage{subfigure}
\usepackage{textcomp}
\usepackage{xcolor}   
\usepackage{braket}
\usepackage{hyperref}
\usepackage{times}

\definecolor{newco}{rgb}{0.89, 0.6, 0.95}

\begin{document}

\title{Designing spin channel geometries for entanglement distribution}
\author{E. K. Levi}
\affiliation{SUPA, School of Physics and Astronomy, University of St Andrews, KY16 9SS, UK}
\author{P. G. Kirton}
\affiliation{SUPA, School of Physics and Astronomy, University of St Andrews, KY16 9SS, UK}
\author{B. W. Lovett}
\affiliation{SUPA, School of Physics and Astronomy, University of St Andrews, KY16 9SS, UK}
\begin{abstract}
We investigate different geometries of spin-1/2 nitrogen impurity channels for distributing entanglement between pairs of remote nitrogen vacancy centers (NVs) in diamond.
To go beyond the system size limits imposed by directly solving the master equation, we implement a matrix product operator method to describe the open system dynamics. In so doing, we provide an early demonstration of how this technique can be used for simulating real systems.
For a fixed NV separation there is an interplay between incoherent impurity spin decay and coherent entanglement transfer: Long transfer time, few-spin systems experience strong dephasing that can be overcome by increasing the number of spins in the channel.
We examine how missing spins and disorder in the coupling strengths affect the dynamics, finding that in some regimes a spin ladder is a more effective conduit for information than a single spin chain. 
\end{abstract}


\maketitle

Nitrogen vacancy centers (NVs) in diamond provide one of the most promising routes for interfacing optics with solid state systems~\cite{Gao2015} because of their long electron and nuclear spin decoherence times that persist to room temperature~\cite{Rondin2014,Balasubramanian2009}.  
However, using coupled NVs in a quantum register requires individual optical addressing, and this sets a minimum spacing that means their direct coupling is almost negligible. 
Such scaling issues could be overcome by using a dark spin channel~\cite{Yao2011}. 
Direct numerical simulations of such multi-spin systems are severely limited by the problem of an exponential growth of Hilbert space with system size. 
However, we show that sophisticated numerical techniques based on matrix product state (MPS) methods are able to overcome such limitations, allowing us to perform numerically exact simulations of systems with up to $\sim27$ spins. 
This allows us to compare different geometries of spin channel, so aiding the design of such systems.  

Fabrication of nitrogen-doped diamond spin-wire structures can be achieved through nitrogen ion implantation followed by an annealing stage to convert some of the nitrogen impurities to NVs~\cite{Yamamoto2013,Antonov2014}. 
The conversion process is not perfectly efficient, but the unconverted impurities can be used as a spin channel~\cite{Cappellaro2011}. 
NVs are amenable to precise measurement and manipulation~\cite{Childress2006}, which has led to an experimentally realizable set of universal quantum operations~\cite{Yao2012,Cappellaro2009}.
There is a large degree of flexibility in the design of the dark nitrogen spin channel geometry that can be used to connect remote NVs. 
It is then of critical importance to learn how resilient different geometries are to missing impurities, and to a distribution of couplings due to imprecise positioning.

Initially it was suggested that spin-1/2 chains provide an ideal method for quantum state transfer (QST)~\cite{Yao2011, Yao2013,Kay2010,Bose2007} and a variety of geometries were explored~\cite{Burgarth2005,Man2014,Burgarth2005a}. 
However, the losses generated by the coupling to the environment of such spin channels mean that it is impractical to use them to directly transfer quantum states between neighboring NVs~\cite{Ping2013}. 
There are ways to circumvent this drawback~\cite{Streltsov2015}, for example it is possible for two distant systems to be entangled via a separable ancilla~\cite{Cubitt2003,Mista2008} and this has been achieved experimentally using single photons~\cite{Fedrizzi2013} and Gaussian beams~\cite{Peuntinger2013,Vollmer2013}. 

We focus here on the alternative option of using \textit{entanglement distillation}~\cite{Bennett1996,Bennett1996a,Bennett1996b}: A large ensemble of weakly entangled pairs are distributed and through local operations and classical communication are refined into a small ensemble of highly entangled pairs -- and then teleportation can be used for state transfer. 
Distribution of entanglement along spin chains has been widely studied~\cite{Apollaro2013,Subrahmanyam2004,Sarkar2011,Ping2013} and very recently extended to dual-rail configurations~\cite{Ji2015,Man2014a}.
 
 \begin{figure}
\includegraphics[scale=0.292,keepaspectratio=true]{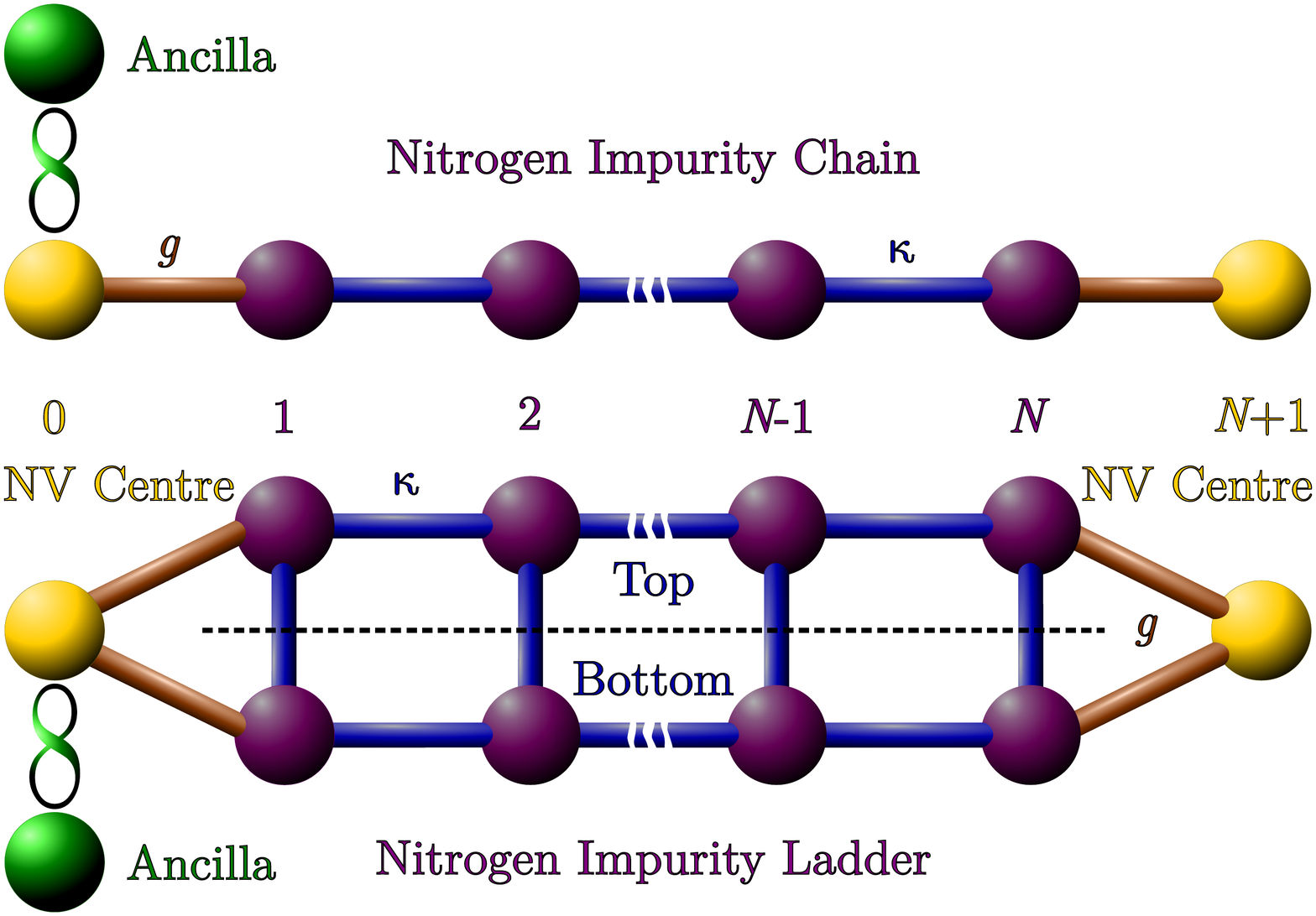}
\caption{\label{fig:CL_Schem}(color online) Schematic diagram of $N$-site nitrogen impurity chain and ladder channels connecting left and right NVs with the indicated couplings. 
  An ancilla spin is initially entangled with the left NV. 
  Intra-channel couplings are of strength $\kappa$ and the NV-channel coupling strength is $g$.
  The index for the ladder spins (and operators) has both a site label $i\in\{1\ldots N\}$ and a top or bottom label as indicated.}
\end{figure}

In this Letter we show how an intelligent choice of the geometry of the spin channel can help to overcome limitations in the manufacturing processes, which may leave certain spins missing and/or lead to disorder in spin-spin coupling strengths.
Obtaining the dynamics of excitations in these spin channels allows us to assess how imperfect manufacturing affects different geometries.
A schematic drawing of the two configurations that we compare is shown in Fig.~\ref{fig:CL_Schem}. 
These are the \textit{chain} in which there is a single route connecting adjacent NVs and the \textit{ladder} which provides multiple routes between the centers.

Our model is illustrated schematically, also in Fig.~\ref{fig:CL_Schem}.
The NV spins are at sites $i=0$ and $i=N+1$ while the spin channel that connects them covers sites $i=1$ to  $N$. 
The Hamiltonian can then be written as a sum of three terms corresponding to the NV, dark spin channel and NV-channel interaction. It reads:
\begin{equation}
H=H_{\rm NV} +H_{\rm C} + H_{\text{NV-C}} .
\end{equation}
The NV Hamiltonian is:
\begin{equation}\label{eq:HNV}
H_{\rm NV}=\frac{\varepsilon}{2}\left(\sigma_0^z+\sigma_{N+1}^z\right),
\end{equation} 
where we assume that two of the three spin ground state levels of the NV can be isolated by applying a magnetic field, and so we use Pauli spin operators, $\sigma^z$,
to describe the Zeeman splitting of the two levels $\varepsilon$.
The nitrogen defects in the channel each have electron spin 1/2 and their Hamiltonian $H_{\rm C}=H_{\rm V}+H_{\rm H}$ can be split into vertical and horizontal components
\begin{subequations}\label{eq:HC}
\begin{align} 
H_{\rm V}= & \sum_{i=1}^N\alpha_i\left(\sigma^+_{i,B}\sigma^-_{i,T} + \text{H.c.}\right),\\
H_{\rm H}= & \sum_{i=1}^{N-1}\sum_{j\in \{B,T\}}\kappa_{i,j}\left(\sigma^+_{i,j}\sigma^-_{i+1,j} + \text{H.c.}\right),
\end{align}
\end{subequations}
with spin raising and lowering operators, $\sigma^\pm$. 
The vertical and horizontal intra-channel couplings are of strength $\alpha$ and $\kappa$ respectively.
Finally, the NV-channel interaction is given by
\begin{multline}\label{eq:HNVC}
H_{\text{NV-C}} = \sum_{j\in \{B,T\}}g_{0,j}\left(  \sigma^+_{0}\sigma^-_{1,j}+ \text{H.c.}\right) \\ 
+  g_{N+1,j}\left(\sigma^+_{N+1}\sigma^-_{N,j} + \text{H.c.}\right),
\end{multline}
where the $g$ give the coupling strengths between the NV and the impurity channel.
This provides the full description for a ladder of spins; to treat a chain we simply omit $H_{\rm V}$ and truncate the summation over $j$ in $H_{\text H}$ and $H_{\text{NV-C}} $.

In our simulations the leftmost NV is prepared in a  maximally entangled state with an ancilla: $\ket{\Psi^-}=\tfrac{1}{\sqrt{2}}\left(\ket{\uparrow\downarrow}-\ket{\downarrow\uparrow}\right)$, while all channel spins and the rightmost NV are initialized with spin down.
The system is then propagated in time and the entanglement of formation, $E$, between the ancilla and the final NV is calculated~\cite{Wootters1998}. 
The value  of $E$ corresponds to the number of shared Bell states per copy required to produce a particular ensemble state using only local operations and classical communication~\cite{Kok2010,Eisert1999}.
Our goal is then to maximize $E$ and so allow for the most efficient entanglement distillation process.

Dissipative processes play an essential role in the dynamics of transport through this type of system. 
To model this we include Markovian decay processes to represent the environment of the surrounding crystal~\cite{Ping2013}:
\begin{multline}\label{eq:lind}
\frac{\mathrm{d}\rho(t)}{\mathrm{d}t}=  -i\left[H,\rho\right] +\gamma_{\rm NV}\left(\mathcal{D}[\sigma^x_0]+\mathcal{D}[\sigma^x_{N+1}]\right) \rho \\ 
  + \gamma_{C}\sum_{i=1}^{N}\sum_{j\in\{B,T\}}\mathcal{D}[\sigma^x_{i,j}]\rho,
\end{multline}
where $\mathcal{D}[X]\rho = X\rho X^\dag-\frac{1}{2}\{X^\dag X,\rho\}$ is the usual Lindblad dissipator.
The Lindblad operators describe the dissipation with associated decay rates $\gamma$ for all NV and channel spins (but not the ancilla) and $\rho$ is the density matrix of the full ancilla-NV-impurity system.
It is possible to realize our Hamiltonian experimentally via steps that include a basis rotation $(x,y,z)\rightarrow(z,-y,x)$ (a full derivation of the mapping can be found in the Supplemental Material of Ref.~\cite{Ping2013}). 
This means modeling of physical spin-flip (phase-flip) noise, characterized by a $T_1$ ($T_2$) coherence lifetime, requires $\sigma^z\,(\sigma^x)$ Lindblad operators.
It has been shown, for these spin channel entanglement distribution systems, that $T_2$ processes are the more destructive type of noise~\cite{Ping2013} and so the equation above only contains $\sigma^x$ Lindblad terms which correspond to $T_2$ dephasing.

To accurately simulate the full dissipative dynamics of the system it is necessary to include effects beyond the single-excitation subspace and work with the full Hilbert-space of our Hamiltonian. 
We make use of two computation methods: 1) for small systems, direct solution of the set of differential equations in Eq.~\ref{eq:lind}~\cite{Johansson2013} and 2) for larger systems a matrix product operator (MPO) formulation~\cite{Perotti2005,Schollwoeck2011}.
For $N<5$ a powerful desktop machine can satisfy the memory requirements of the direct solution.
For $N\geq5$ the MPO-based implementation overcomes the exponential memory requirements of the problem, but for $N<5$ this is slower than direct solution.

A general state of the ladder (or chain), including the NVs and ancilla, can be written as
\begin{equation}\label{eq:GMrho}
\rho=\sum_{\{n_0,\ldots n_{N+1}\}}D_{n_0, \ldots n_{N+1}}\sigma_0^{n_0}\otimes\ldots \otimes\sigma_{N+1}^{n_{N+1}},
\end{equation}
where the complete basis describing the density matrix at each site (consisting of two physical spins for the ladder) $\{\sigma_{i}\}$ are the sixteen Gell-Mann matrices, a generalization of the Pauli matrices to a two spin (four level) system~\cite{Sbaih2013}.
The rightmost NV is paired with a dummy, non-interacting spin.
So far this description of the state is exact. The efficiency gain from using an MPO decomposition of Eq.~\ref{eq:GMrho} comes from re-expressing the coefficients as
\begin{multline}\label{eq:mpo}
D_{n_0, \ldots n_{N+1}}=\sum_{\{\nu_k\}}\Gamma^{[0]n_0}_{\nu_0} \lambda^{[0]}_{\nu_0}\Gamma^{[1]n_1}_{\nu_0\nu_1}\lambda^{[1]}_{\nu_1}\times\ldots \\ 
\times  \Gamma^{[N]n_N}_{\nu_{N-1}\nu_N}\lambda^{[N+1]}_{\nu_N}\Gamma^{[N+1]n_{N+1}}_{\nu_N}.
\end{multline}
This procedure is equivalent to performing a Schmidt decomposition at each site, with the $\{\lambda^{[i]}\}$ vectors containing the Schmidt coefficients~\cite{Orus2014}.
Truncation of the bond dimension, $\{\nu_k\}$, then only keeps the most important Schmidt vectors in the description of the system, reducing the scaling with system size to be polynominal. To check convergence we increase the number of singular values retained until the dynamics are insensitive to adding more.
Time evolution of our MPO relies on the time evolving block decimation (TEBD) method~\cite{Vidal2003,Vidal2004} extended for dealing with density matrices~\cite{Zwolak2004,Verstraete2004}. The code used here is modified from already tested TEBD MPO code~\cite{Joshi2013,Schiro2015}.

To understand the effects of both the coherent evolution and lossy dynamics we begin by studying our model when all the couplings are uniform,  i.e.\ by setting $g_{0,j}=g_{N+1,j}=g$ and $\alpha_i=\kappa_{i,j}=\kappa$ in Eq.~\eqref{eq:HC}.
To maximize transfer speed we choose $g=\kappa=\mu_0g_e^2\mu_B^2/8\pi r^3$, where $g_e$ is the electron g-factor and $r$ denotes the spin separation distance. 
In this strong coupling regime the maximum entanglement of formation is largely independent of $\varepsilon$; this is in contrast to the weak coupling $g\ll\kappa$ limit~\cite{Ping2013} where it is necessary to target a particular channel eigenmode. 
We are thus free to choose $\varepsilon=0$. We also fix the NV decay rate $\gamma_{\rm NV}=1/T_2=0.1$~\si{\kilo\hertz}~\cite{Rondin2014}.

\begin{figure}
\includegraphics[scale=0.233,keepaspectratio=true]{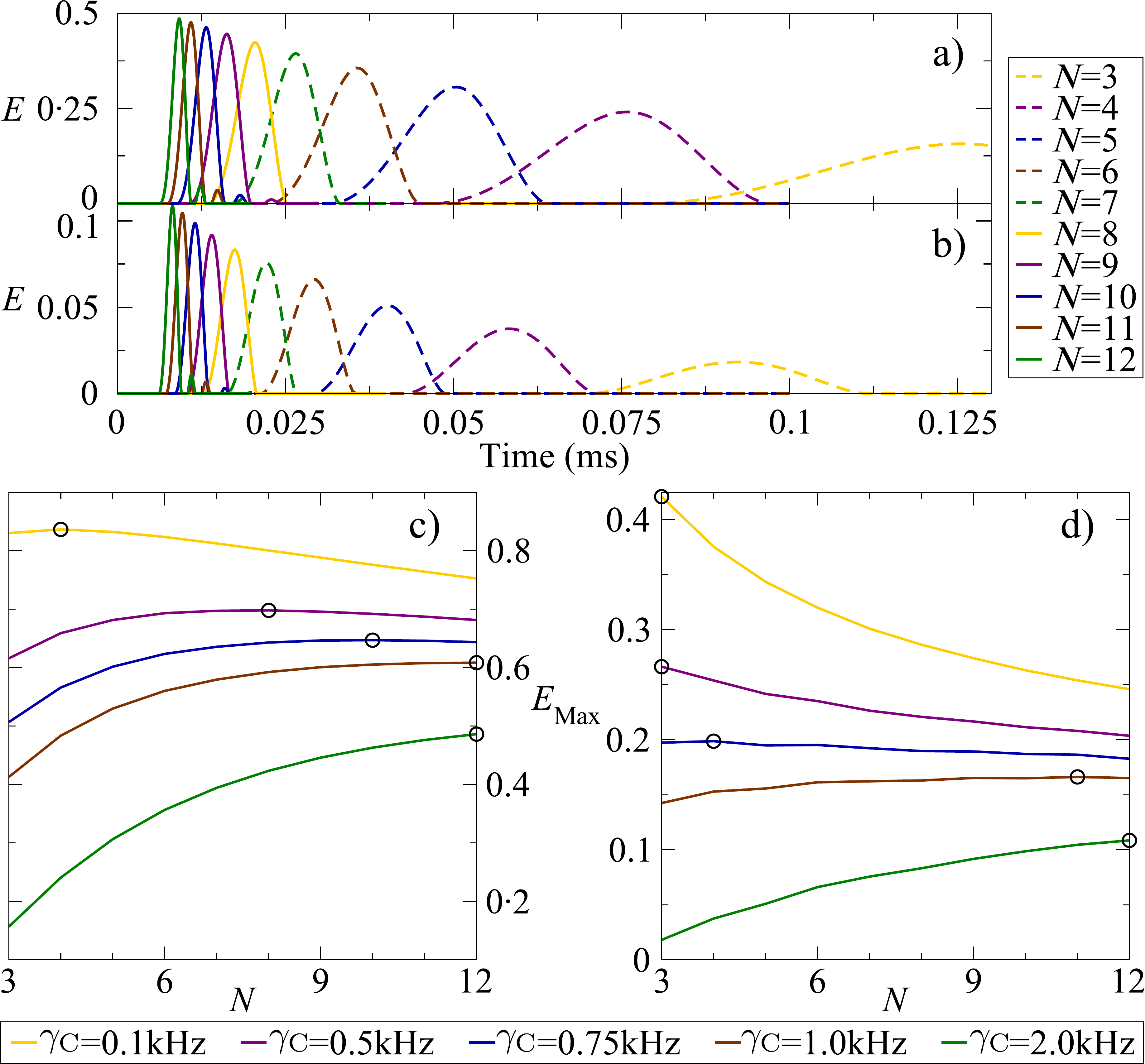}
\caption{\label{fig:40nm}(color online) Dynamics of the entanglement of formation, $E$, for increasing number of spins in the channel $N$ (right to left) in (a) chains and (b) ladders between NVs spaced 40~\si{\nano\metre} apart, and loss rate $\gamma_{\text C}=2$~\si{\kilo\hertz}.
The lower panels show the maximum $E$, for different $\gamma_{\text C}$ values (increasing from top to bottom) as a function of $N$ for (c) chains and (d) ladders. 
The $N$ that maximizes $E_{\text{Max}}$ is circled.}
\end{figure}

In Fig.~\ref{fig:40nm} we show the dynamics of $E$ while increasing the number of spins in the channel with a fixed NV separation of 40~\si{\nano\meter}~\cite{Chen2015,Haeusler2014}.
We limit ourselves to a maximum of $N=12$, since this is approaching to the limit of our numerical capability. For an $N=12$ ladder we are already simulating exact dynamics for 27 spins, a Hilbert space dimension of more than $10^8$; calculating converged dynamics of longer channels requires smaller time steps and increased bond dimension.

The fidelity of transfer is determined by the competition between the coherent transfer rate and the loss of information to the environment. 
As $N$ gets larger, the spins get closer and $g$ and $\kappa$ increase -- as can be seen in Fig.~\ref{fig:40nm}(a) and (b), this expedites the entanglement transfer. 
Competing with this effect is the fact that as $N$ increases the number of loss channels also increases. 
For the value of $\gamma_{\rm C}=2$~\si{\kilo\hertz} in Fig.~\ref{fig:40nm}(a) and (b), the long transfer time for smaller numbers of spins in the channel is clearly seen to be the limiting factor, rather than the effect of fewer spins undergoing decay. 
However, as can be seen in Fig.~\ref{fig:40nm}(c) and (d) the optimal $N$ becomes larger for increasing $\gamma_{\rm C}$; when $\gamma_{\rm C}$ is small the system is able to remain efficient even with a slow transfer rate, but as $\gamma_{\rm C}$ increases the faster transfer through a longer channel means the decay is less important.

The value of $E$ for ladders in Fig.~\ref{fig:40nm}(b) and (d) is always lower than for chains of the same length. 
This is because the ladder is constructed from more spins than the chain and so always has a larger total effective decay rate, but the extra spins in the ladder allow faster transfer of entanglement.

Let us next investigate the possibility that spins may be missing from a manufactured channel. In what follows we will fix the value of $\gamma_{\rm C}=2$~\si{\kilo\hertz} as this will allow us to clearly show the physics of interest in an experimentally motivated parameter regime~\cite{Takahashi2008,Bar-Gill2013}.
As seen in Fig.~\ref{fig:40nm}, at this decay rate the coupling strength associated with $N=12$ provides the maximum fidelity of entanglement distribution so we fix the interspin separation at $r=40/13$~\si{\nano\meter} and hence $g=\kappa\sim 0.9$~\si{\mega\hertz}.
Therefore the total length of the channel now increases as we add spins to it.

\begin{figure}
\includegraphics[scale=0.303,keepaspectratio=true]{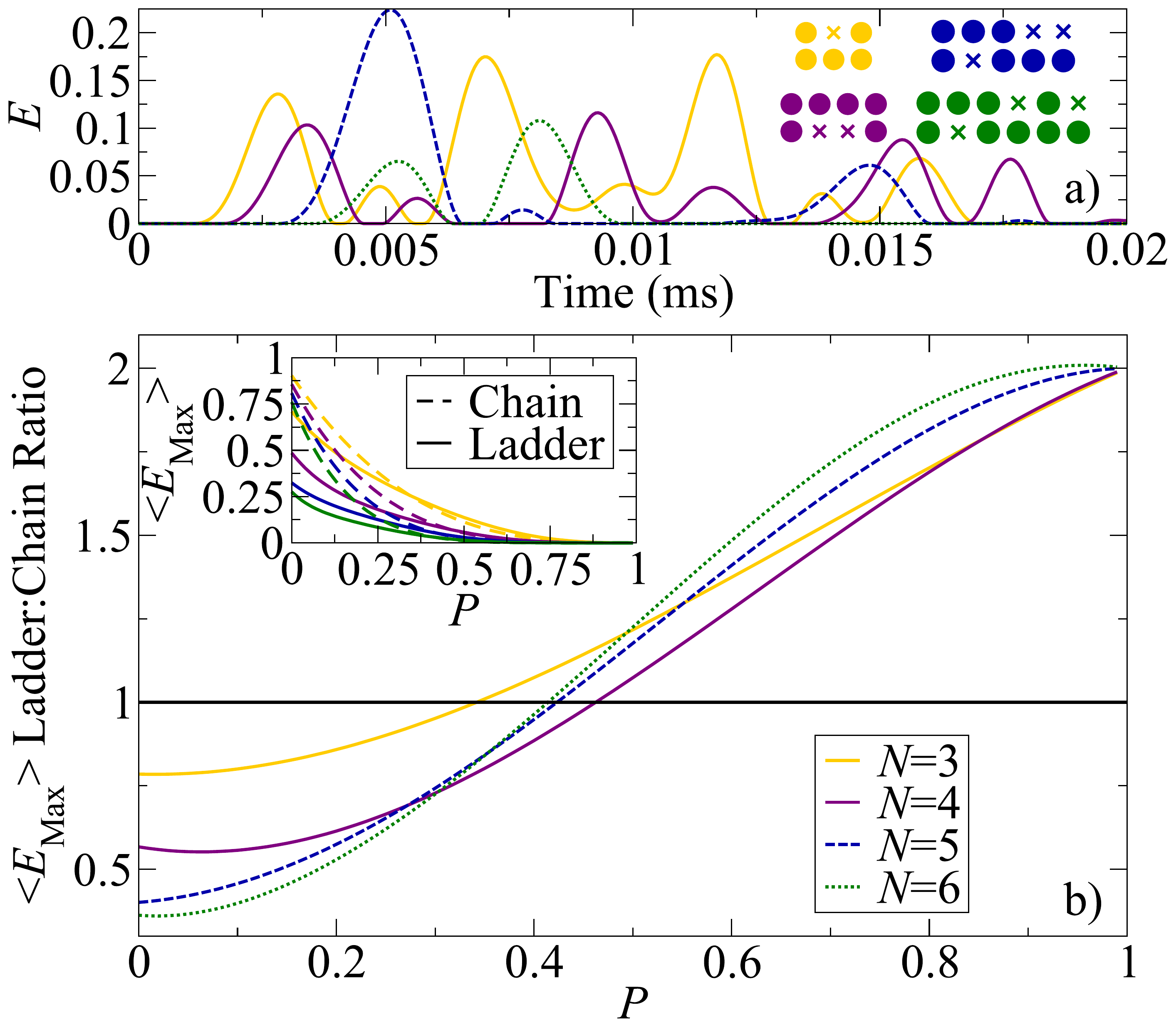}
\caption{\label{fig:missing}(color online) a) Dynamics of $E$ for various impurity ladder configurations  with missing spins. 
			     These are schematically illustrated in the inset. 
			     The crosses denote missing spins.
			     b) The ratio of the values of $\langle E_{\mathrm{Max}}\rangle$ for the ladder and chain as a function of the probability for each spin to be missing $P$ for a variety of channel lengths $N$. 
			     The inset shows the actual values of $\langle E_{\mathrm{Max}}\rangle$ with increasing channel length from top to bottom.}
\end{figure}

It is immediately obvious that a single spin missing from a nearest-neighbor interacting chain constitutes a catastrophic break rendering entanglement distribution impossible, but as can be seen from the example results for various missing spin configurations in Fig.~\ref{fig:missing}(a), this is not the case for a ladder.
To investigate this further we look at how the average efficiency of a channel decays as the probability that a given spin is missing, $P$, increases. We calculate the average maximum entanglement of formation through
\begin{equation}
\langle E_{\mathrm{Max}}\rangle(P)=\sum_{\mathrm{c}}P^{m_\mathrm{c}}(1-P)^{M-m_{\mathrm{c}}}E_{\mathrm{Max,c}}
\end{equation}
where the summation is over all possible missing spin configurations, $M$ is the total number of spins and $m_\mathrm{c}$ is the number of missing spins in the configuration ${\rm c}$. 

The value of $\langle E_{\mathrm{Max}}\rangle$ for channels of length $N=3$ to 6 can be seen inset to Fig~\ref{fig:missing}(b). The dependence is intuitive: a higher $P$ causes a reduction in $\langle E_{\mathrm{Max}}\rangle$ for both chain and ladder.
To see more clearly which channel type performs best, the main panel of Fig.~\ref{fig:missing}(b) shows the ladder:chain ratio of $\langle E_{\mathrm{Max}}\rangle$ for each $N$.
It is clear a ladder is more robust to missing spins, but because there are more possible spins to lose in a ladder, it is not until a relatively large value of $P$ that using a ladder becomes beneficial. 
As the channel length is increased, the trend (except for in the very short $N=3$ case) is that the ladder starts to outperform the chain at a lower value of $P$. 
We expect this trend to continue to larger values of $N$; in a long chain even a very small value of $P$ will cause catastrophic failures to dominate $\langle E_{\mathrm{Max}}\rangle$, but we cannot verify this since we are constrained by computational resources: The number of configurations which need to be simulated grows very quickly with the length of the channel.

Placing spins in a chain or ladder configuration naturally has some inherent fabrication imprecision. This has the direct consequence that the couplings between spins will take on some distribution of values. 
We investigate this effect by introducing random couplings about the ideal $\kappa(r=40/13$~\si{\nano\meter}$)$ value. 
We need to choose a distribution which vanishes at $\kappa=0$ and has a long tail at large $\kappa$ to approximately describe the effects of randomly placing spins around some mean value, with this in mind we use a log-normal distribution~\cite{Johnson1994}. 
The number of disorder realizations necessary for numerics to converge to good accuracy means that we are limited to studying channels with $N=3$ and 4.

\begin{figure}
\includegraphics[scale=0.305,keepaspectratio=true]{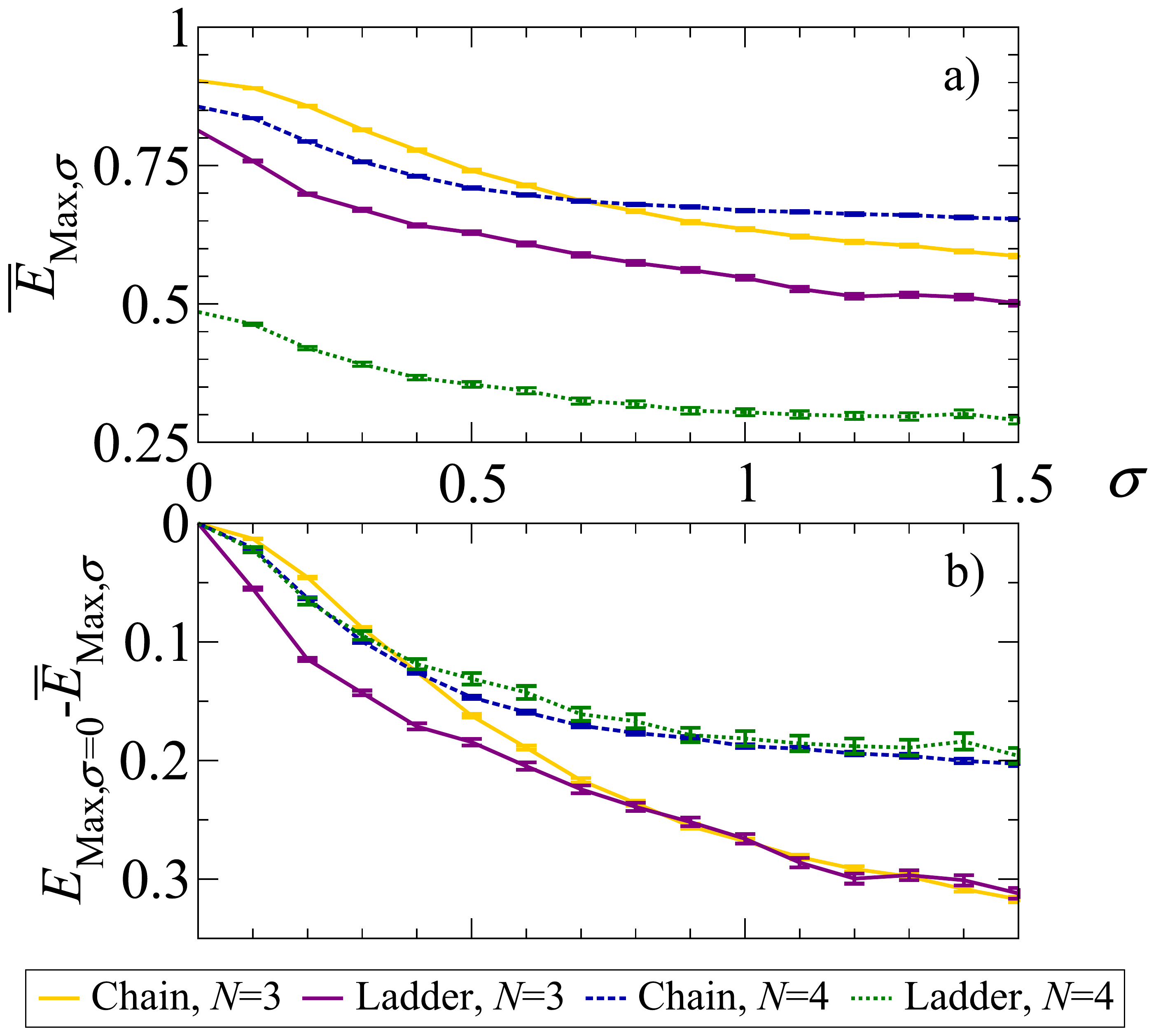} 
\caption{\label{fig:lognorm} (color online). a) Disorder averaged $\overline{E_{\mathrm{Max},\sigma}}$ for randomly assigned intra-channel bond strengths from a log-normal distribution with standard deviation $\sigma$. 
			     The error bars are calculated using the standard error of the mean.
			     b) The difference between the ideal $\sigma=0$ case and its disordered equivalent. Each point was averaged over $k$ disorder realisations where $k=12000$ for the $N=3$ chain, $k=10000$ for the $N=4$ chain, $k=2000$ for the $N=3$ ladder and $k=600$ for the $N=4$ ladder. }
\end{figure}

In Fig.~\ref{fig:lognorm} we show the disorder averaged maximum entanglement of formation, $\overline{E_{\mathrm{Max},\sigma}}$ as a function of the standard deviation, $\sigma$, of the distribution.
We find that a broader distribution leads to a lower $\overline{E_{\mathrm{Max},\sigma}}$ for both chain and ladder, but in this case the chain always outperforms the ladder for all tested $\sigma$.
As can be seen in Fig.~\ref{fig:lognorm}(b), the deviation from a perfect channel of a given length is very similar for both chains and ladders.
It is also clear from these results that although a longer channel severely limits the distribution fidelity of the ladder, it also reduces the relative overall effect of the disorder.

In conclusion, using matrix product operators to perform numerically exact quantum simulations in 
much larger Hilbert spaces than is feasible for direct solution has enabled us to study the dynamics of many-spin channels in an open environment. 
The types of numerical techniques used here are applicable to studying the behavior of a wide variety of similar systems. 
For example it could be applied to conduction in quasi-1D channels such as in carbon nanotubes, polymers or DNA~\cite{Tans1997,Fink1999}. 
Generalizations to higher dimensions are possible through projected entangled pair states (PEPS)~\cite{Orus2014}.

We have been able analyze the benefits and drawbacks of using different geometries of spin channels to distribute entanglement between two separated NVs. 
We find that in the ideal case, with no manufacturing imperfections, using simple chains is optimal. 
When spins are missing ladders perform better and with intra-channel disorder both geometries scale similarly.
Extrapolating our results we believe that a study combining both of these should find that ladders outperform chains after a similar threshold as shown for our missing spin results. 
Unfortunately, numerical limitations make it impossible to verify this directly.

An interesting next step would be to examine what happens when there are interactions beyond nearest neighbor coupling.
Whilst this should make both channels more robust against defects it would also allow chains to continue to function when a spin is missing, causing them to be more robust to this kind of defect.
Ladders and chains both have strengths and drawbacks when used for entanglement distribution. The particular kinds and scales of dissipation and disorder determine which is the best geometry to use.

\acknowledgments{We thank J.~Keeling for providing the initial MPS code used in this study, and for suggesting that it could be used for modeling spin ladders. 
EKL acknowledges support from EPSRC (EP/G03673X/1).  
PGK acknowledges support from EPSRC grant EP/M010910/1.}

\bibliographystyle{apsrev4-1}
\bibliography{NV-paper}{}
\end{document}